\documentclass[twocolumn,showpacs,preprintnumbers,amsmath,amssymb]{revtex4}

\usepackage{graphicx}
\usepackage{dcolumn}
\usepackage{bm}

\begin{document}

\preprint{}

\title{Photonic spin Hall effect in topological insulators}
\author{Xinxing Zhou}
\author{Jin Zhang}
\author{Xiaohui Ling}
\author{Shizhen Chen}
{}
\author{Hailu Luo}\email{hailuluo@hnu.edu.cn}
\author{Shuangchun Wen}\email{scwen@hnu.edu.cn}
\affiliation{Key Laboratory for Micro-/Nano-Optoelectronic Devices
of Ministry of Education, College of Physics and Microelectronic
Science, Hunan University, Changsha 410082, People's Republic of
China}
\date{\today}

\begin{abstract}
In this paper we theoretically investigate the photonic spin Hall
effect (SHE) of a Gaussian beam reflected from the interface between
air and topological insulators (TIs). The photonic SHE is attributed
to spin-orbit coupling and manifests itself as in-plane and
transverse spin-dependent splitting. We reveal that the spin-orbit
coupling effect in TIs can be routed by adjusting the axion angle
variations. Unlike the transverse spin-dependent splitting, we find
that the in-plane one is sensitive to the axion angle. It is shown
that the polarization structure in magneto-optical Kerr effect is
significantly altered due to the spin-dependent splitting in
photonic SHE. We theoretically propose a weak measurement method to
determine the strength of axion coupling by probing the in-plane
splitting of photonic SHE.
\end{abstract}


\maketitle

\section{Introduction}\label{SecI}
The spin Hall effect (SHE) is a transport effect of inducing
transverse spin currents perpendicular to the applied electric field
direction~\cite{Hirsch1999,Kato2004}. The SHE offers a promising
method for spatially separating electron spins which could be
applied to develop new spintronic devices~\cite{Sih2005}. It is well
known that the physical mechanism of SHE in electric systems is
attributed to the spin-orbit coupling. Recently, a material called
topological insulators (TIs) has aroused tremendous
interest~\cite{Qi2010,Moore2009}. It has gapless helical surface
states owing to the topological protection of the time-reversal
symmetry and represents a full energy gap in the
bulk~\cite{Fu2007,Maciejko2010,Chang2009}. It is found that an
amazing effect named quantum SHE can exist in TIs due to the
spin-orbit coupling. Here the spin up and spin down electrons move
in opposite directions on the surface edge of TIs with no external
magnetic field~\cite{Bernevig2006,Qi2011}. The quantum SHE holds
great potential applications especially for resulting in new
spintronic or magnetoelectric devices.

The photonic SHE is photonic version of the SHE in electronic
systems, in which the spin photons play the role of the spin
charges, and a refractive index gradient plays the role of applied
electric field~\cite{Onoda2004,Bliokh2006,Hosten2008}. This effect
manifests itself as in-plane and transverse spin-dependent splitting
of left- and right-circular components when a spatially confined
light beam is reflected or transmitted at an interface. The photonic
SHE is currently attracting growing attention and has been
intensively investigated in different physical systems such as
optical physics~\cite{Bliokh2008,Aiello2008,Luo2009,Hermosa2011},
high-energy physics~\cite{Gosselin2007,Dartora2011}, semiconductor
physics~\cite{Menard2009,Menard2010}, and
plasmonics~\cite{Shitrit2011,Gorodetski2012,Shitrit2013}. The
photonic SHE is generally believed to be a result of an effective
spin-orbit coupling that is related to the Berry phase.

In this work, similar to the electron SHE in TIs with spin-orbit
coupling, we evolve the photonic SHE in TIs with spin-orbit coupling
that is an interaction of photon spin and the trajectory of beam
propagation. The electromagnetic effect called axion coupling can
occur in TIs when weak time-reversal breaking perturbation is
introduced~\cite{Maciejko2010,Chang2009}. This coupling is described
by the effective Lagrangian density $\mathcal {L}_{\Theta}=(\alpha
\Theta/4\pi^{2})\mathbf{E}\cdot\mathbf{B}$. Here,
$\alpha=e^{2}/\hbar c$ is the fine-structure constant and $\Theta$
is the axion angle corresponding to the axion coupling effect. Under
this condition, an applied electric field can induce magnetization
and a magnetic can cause polarization. We revealed that, due to the
axion coupling effect of TIs, the spin-orbit coupling in photonic
SHE can be modulated by adjusting the axion angle. The field
distribution of the reflected beam is analyzed and we find that the
polarization structure in magneto-optical Kerr effect is
significantly altered due to the spin-dependent splitting in
photonic SHE.

The rest of the paper is organized as follows. In Sec. II, we
establish a general theoretical model to describe the photonic SHE
in TIs. Here the quantitative relationship between the magnitude of
axion angle, spin-dependent shifts and Kerr rotation angle are
established. Next, according to the numerical calculation, we reveal
that the in-plane spin-dependent splitting can be modulated by
adjusting the axion angle and can occur in the case of horizontal
and vertical polarizations showing difference from the usual
air-glass interface~\cite{Qin2011}. However, the transverse
spin-dependent splitting is insensitive to the axion coupling
variations for a certain range. Additionally a signal enhancement
technique known as weak measurements is theoretically proposed to
detect this topological phenomenon. Finally, a conclusion is given
in Sec. IV.
\begin{figure}
\includegraphics[width=8.5cm]{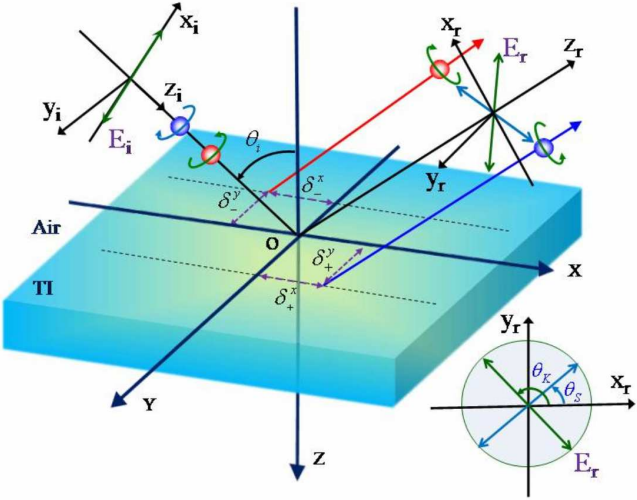}%
\caption{\label{Fig1} (Color online) A linearly polarized Gaussian
beam reflected on the interface between air and TIs occurs photonic
SHE and magneto-optical Kerr effect. Here, $\theta_{i}$ is the
incident angle. $\delta_{\pm}^{x}$ and $\delta_{\pm}^{y}$ indicate
the in-plane (in x direction) and transverse (in y direction)
displacements of left- or right-circularly polarized component. The
inset: $\theta_{K}$ stands for the Kerr rotation angle and
$\theta_{S}$ is the spin-dependent splitting rotation angle. We
consider the relatively simple external perturbation with magnetic
field. }
\end{figure}

\section{Theoretical analysis}\label{SecII}
Figure~\ref{Fig1} schematically illustrates a polarized Gaussian
beam reflection at an air-TIs interface. When the axion coupling
effect appears, the normal rules of electromagnetic wave propagation
are modified~\cite{Maciejko2010}. As a result, the magneto-optical
Kerr effect manifesting for the polarization plane of reflected
light acquiring a certain rotation occurs. Importantly, the photonic
SHE manifesting for the left- and right-circularly polarized
components obtaining opposite shifts can also happen, which is
ignored by the previous Kerr effect research. We first theoretically
establish the relationship between axion coupling and these two
physical phenomena. In this paper, we only consider the incident
light beam with horizontal (H) polarization. The vertical (V)
polarization condition can be analyzed in the similar way. The
electromagnetic response of a system under the condition of axion
coupling takes the usual Maxwell's equations. We just need to modify
the electric displacement $\mathbf{D}$ and magnetic field
$\mathbf{H}$ as: $\mathbf{D}=\epsilon \mathbf{E}
+\alpha(\Theta/\pi)\mathbf{B}$ and $\mathbf{H}=(\mathbf{B}/\mu)
-\alpha(\Theta/\pi)\mathbf{E}$~\cite{Chang2009}. Here
$\alpha=e^{2}/\hbar c$ is the fine-structure constant and $\Theta$
is the axion angle corresponding to the axion coupling effect which
can be manipulated by the external perturbation with a thin magnetic
layer or applying an external static magnetic field perpendicular to
the surface. Here we only discuss the relatively simple condition of
external magnetic field.

We first consider the electromagnetic plane waves reflection on the
air-TIs interface. One can see that the boundary conditions of the
interface with no surface charge and surface current take the usual
forms. So the normal components of $\mathbf{D}$ and $\mathbf{B}$ and
the tangential components of $\mathbf{E}$ and $\mathbf{H}$ are also
continuous across the interface. With the help of the boundary
conditions, we can obtain the $2\times2$ Fresnel reflection matrix
describing the relations between the incident and the reflection
fields (as for the plane wave)~\cite{Chang2009}:
\begin{eqnarray}
\frac{1}{\Lambda}\left[
\begin{array}{cc}
(1-n^{2}-\bar{\alpha}^{2})+n\xi_{-} & 2\bar{\alpha}\\
2\bar{\alpha} & -(1-n^{2}-\bar{\alpha}^{2})+n\xi_{-}
\end{array}\right] \label{TIRM}.
\end{eqnarray}
Here $\bar{\alpha}=\alpha\Theta/\pi$,
$\Lambda=1+n^{2}+\bar{\alpha}^{2}+n\xi_{+}$ and
$\xi_{\pm}=(\cos\theta_{i}/\cos\theta_{t})\pm(\cos\theta_{t}/\cos\theta_{i})$.
The cross-polarized reflection coefficients stem from the term
$\bar{\alpha}$ corresponding to the axion coupling, which should be
distinguished from the chiral metamaterials~\cite{Wang2011,Xu2011}.
In our work, we focus our attention on the photonic SHE and Kerr
rotation by considering the incident beam with H polarization. If
the axion coupling effect vanishes, the cross-polarized reflection
coefficients obtaining from the term $\bar{\alpha}$ turn to zero and
there is no Kerr rotation. And the reflection matrix can reduce to
the normal style~\cite{Zhou2012a}.

To simplify the following calculation, we can rewrite the reflection
matrix from Eq.~(\ref{TIRM}) as
\begin{eqnarray}
\left[
\begin{array}{cc}
r_{pp} & r_{ps}\\
r_{sp} & r_{ss}
\end{array}\right] \label{SRM}.
\end{eqnarray}
For incident light linearly polarized in the x direction or y
direction, the Kerr rotation angle are defined by
$\tan\theta_{K}=E_{r}^{y}/E_{r}^{x}$. We can see that the Kerr
effect manifests itself as the rotation angle of polarization
direction of reflected light's central wave vector. So the Kerr
rotation angle can be obtained by considering the incident light
beam with the Jones vector $(1,0)^{T}$ or $(0,1)^{T}$ standing for
the H or V polarization. The results can be written as
$\tan\theta_{K}=r_{sp}/r_{pp}$ or $\tan\theta_{K}=r_{ss}/r_{ps}$ for
H or V polarization incident. We note that there is no Kerr rotation
when the cross-polarized reflection coefficients $r_{sp}$ and
$r_{ps}$ are equal to zero. So the Kerr rotation discussed here
origins from the axion coupling effect.

Next we study the photonic SHE and calculate the corresponding
displacements of in-plane and transverse spin-dependent splitting,
respectively. Let us first consider the paraxial beam reflection on
the TIs interface. In the spin basis set, the angular spectrum can
be written as
$\tilde{\mathbf{E}}_i^H=(\tilde{\mathbf{E}}_{i+}+\tilde{\mathbf{E}}_{i-})/{\sqrt{2}}$
and
$\tilde{\mathbf{E}}_i^V=i(\tilde{\mathbf{E}}_{i-}-\tilde{\mathbf{E}}_{i+})/{\sqrt{2}}$.
The positive and negative signs in the supscripts denote the left-
and right-circularly polarized (spin) components. In paraxial
optics, the incident field of a localized wave packet whose spectrum
is arbitrarily narrow, can be written as
\begin{equation}
\widetilde{\mathbf{E}}_{i\pm}=(\mathbf{e}_{ix}+i\sigma\mathbf{e}_{iy})\frac{w_{0}}{\sqrt{2\pi}}\exp
\left[-\frac{w_{0}^{2}(k_{ix}^{2}+k_{iy}^{2})}{4}\right]\label{IEF},
\end{equation}
where $w_{0}$ is the beam waist. The polarization operator
$\sigma=\pm1$ corresponds to left- and right-circularly polarized
light.

To accurately describe the in-plane and transverse spin-dependent
splitting, we need to determine the reflection of arbitrary
wave-vector components. After the coordinate rotation and combining
with Eq.~(\ref{SRM}), we can obtain the reflected angular spectrum
by means of the relation
$\tilde{\mathbf{E}_r}(k_{rx},k_{ry})=\mathbf{M}_{R}\tilde{\mathbf{E}_i}(k_{ix},k_{iy})$~\cite{Luo2011}.
Here $\mathbf{M}_{R}$ stands for
\begin{eqnarray}
\left[
\begin{array}{cc}
r_{pp}+\frac{k_{ry}\cot\theta_{i}(r_{sp}-r_{ps})}{k_{0}} &r_{ps}+\frac{k_{ry}\cot\theta_{i}(r_{pp}-r_{ss})}{k_{0}} \\
r_{sp}-\frac{k_{ry}\cot\theta_{i}(r_{pp}-r_{ss})}{k_{0}} &
r_{ss}+\frac{k_{ry}\cot\theta_{i}(r_{sp}-r_{ps})}{k_{0}}
\end{array}\right]\label{FRM}.
\end{eqnarray}

In the above equations, $r_{pp}$, $r_{ps}$, $r_{sp}$ and $r_{ss}$
denote Fresnel reflection coefficients according from
Eq.~(\ref{SRM}). $k_{0}$ is the wave number in free space. By making
use of a Taylor series expansion based on the arbitrary angular
spectrum component, we expand the Fresnel reflection coefficients
around the central wave vector for considering the tiny in-plane
spread of wave-vectors:
\begin{eqnarray}
r_{ab}=r_{ab}(\theta_{i})+\frac{\partial r_{ab}}{\partial
\theta_{i}}\frac{k_{ix}}{k_{0}}. \label{Taylor}
\end{eqnarray}
Here $r_{ab}$ denotes the Fresnel reflection coefficients where an
incident wave with $b$ polarization accompanied by a reflected wave
with $a$ polarization (a and b stand for either s or p component),
$r_{ab}(\theta_{i})$ corresponds to the central wave vector at the
incident angle $\theta_{i}$.

Putting everything together, we can obtain the reflected angular
spectrum of the practical polarized Gaussian beam in the case of H
polarization:
\begin{eqnarray}
\widetilde{\mathbf{E}}_{r} &=&
\frac{r_{pp}(\theta_{i})}{\sqrt{2}}\bigg\{\exp\left[k_{rx}(+i\eta-\vartheta)\right]\exp(+ik_{ry}\delta) \nonumber\\
&&-i\frac{r_{sp}(\theta_{i})}{r_{pp}(\theta_{i})}\bigg\}\widetilde{\mathbf{E}}_{r+}
+\frac{r_{pp}(\theta_{i})}{\sqrt{2}}\bigg\{\exp\left[k_{rx}(-i\eta-\vartheta)\right]\nonumber\\
&&\times\exp(-ik_{ry}\delta)
+i\frac{r_{sp}(\theta_{i})}{r_{pp}(\theta_{i})}\bigg\}\widetilde{\mathbf{E}}_{r-}\label{spectrum},
\end{eqnarray}
provided that $k_{rx}\eta$$\ll$1, $k_{rx}\vartheta$$\ll$1 and
$k_{ry}\delta$$\ll$1. In the above equation $\eta$=$(\partial
r_{sp}/\partial \theta_{i})/r_{pp}(\theta_{i})k_{0}$,
$\vartheta$=$(\partial r_{pp}/\partial
\theta_{i})/r_{pp}(\theta_{i})k_{0}$ and
$\delta$=$[1+r_{ss}(\theta_{i})/r_{pp}(\theta_{i})]\cot
\theta_{i}/k_{0}$. The terms $\exp(\pm ik_{rx}\eta)$ and $\exp(\pm
ik_{ry}\delta)$ represent the spin-orbit coupling. Here we can see
that the spin-orbit coupling related items $\eta$ and $\delta$ are
affected by the axion coupling effect. It is well known that the
photonic SHE manifests itself as the spin-orbit coupling. So the
in-plane and transverse spin-dependent splitting can be modulated
through the axion coupling effect. Additionally, the other electric
field components $\exp(-ik_{rx}\vartheta)$ and
$ir_{sp}(\theta_{i})/r_{pp}(\theta_{i})$ are not the spin-orbit
coupling terms, however, they can affect the spin-dependent
splitting. Especially, the term
$ir_{sp}(\theta_{i})/r_{pp}(\theta_{i})$ also plays the great role
in Kerr rotation. We note that the correction proportional to wave
vector component $k_{ix}$ ignored by the previous work is
responsible for the in-plane spin-dependent
splitting~\cite{Qin2011}.

The photonic SHE is described for the left- and right-circularly
polarized components undergoing the in-plane and transverse shifts.
So the reflected field centroid should be determined. At any given
plane $z_a=\text{const.}$, the displacements of field centroid
compared to the geometrical-optics prediction is given by
\begin{equation}
\delta_{\pm}^{x,y}= \frac{\int\int \tilde{\mathbf{E}}^{\ast}
i\partial_{\mathbf{k}_{\bot}}\tilde{\mathbf{E}} dk_{rx}
dk_{ry}}{\int\int \tilde{\mathbf{E}}^{\ast}\tilde{\mathbf{E}}
dk_{rx} dk_{ry}}\label{centroid}.
\end{equation}
Here
$\delta_{\pm}^{x,y}=\delta_{\pm}^{x}\mathbf{e}_{rx}+\delta_{\pm}^{y}\mathbf{e}_{ry}$
and $\partial_{\mathbf{k_\perp}}=\frac{\partial}{\partial
k_{rx}}\mathbf{e}_{rx}+\frac{\partial}{\partial
k_{ry}}\mathbf{e}_{ry}$. Substituting Eq.~(\ref{spectrum}) into
Eq.~(\ref{centroid}), we can get the in-plane and transverse
spin-dependent displacements. It should be noted that, according to
Eq.~(\ref{spectrum}), we only consider the electric field component
$\exp\left[k_{rx}(\pm i\eta-\vartheta)\right]\exp(\pm
ik_{ry}\delta)$ including the spin-orbit coupling part for field
centroid calculation. Hence the in-plane and transverse
spin-dependent shifts in photonic SHE can be obtained.

\section{Results and discussion}\label{SecIII}
When a linearly polarized light reflects from a medium surrounded by
the magnetic field, the polarization plane undergoes a rotation is
the magneto-optical Kerr effect~\cite{Katayama1988,Tse2010}. The
Kerr effect can be used for determining the surface sensitivity of
material, studying magnetic materials on account of its robustness
and so on~\cite{Buchmeier2009}. Here we investigate the Kerr
rotation angles changing with different axion coupling under the
condition of external magnetic field. We only consider the incident
light beam with H polarization. The direction of Kerr rotation is
referred to the polarization direction of reflected light's central
wave vector. Generally speaking, under the condition of H
polarization, the polarization direction of reflected light's
central wave vector acquires no rotation for the absence of axion
coupling. So there is no Kerr effect. When time reversal breaking
perturbation is introduced on the TIs surface, the material becomes
fully gapped and the axion coupling effect happens, which induces
the Kerr effect.

\begin{figure}
\includegraphics[width=8cm]{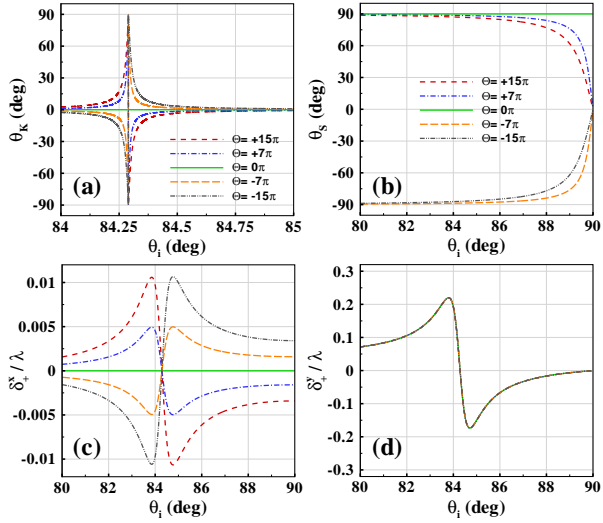}
\caption{\label{Fig2} (Color online) The photonic SHE and
magneto-optical Kerr effect induced by axion coupling at air-TIs
interfaces in the case of H polarization. Here the parameters are
the refractive index of TIs n=10 (appropriate for the TIs such as
$Bi_{1-x}Se_{x}$). The beam waist is selected as $w_{0}=20\lambda$.
(a) shows the Kerr rotation angle with different axion angles. (b)
denotes the spin-dependent splitting rotation angle which satisfies
the relationship $\tan\theta_{S}=\delta_{+}^{y}/\delta_{+}^{x}$. (c)
and (d) describe the in-plane and transverse displacements,
respectively. }
\end{figure}

The axion angle $\Theta$ is shown to be quantized in odd integer
values of $\pi$: $\Theta=(2n+1)\pi$, where $n\in\mathbb{Z}$. Here we
choose the axion angle as $\Theta=0, \pm7\pi, \pm15\pi$. And the
refractive index of the TIs is chosen as n=10 (appropriate for the
TIs such as $Bi_{1-x}Se_{x}$)~\cite{Chang2009}. The beam waist is
selected as $w_{0}=20\lambda$. Figure~\ref{Fig2}(a) shows the Kerr
rotation angle $\theta_{K}$ versus the incident angle $\theta_{i}$
for H polarization. Since the $\theta_{K}$ represents a tiny values
for $\theta_{i}$ in the range of $0^{\circ}$ to $84^{\circ}$ and
$85^{\circ}$ to $90^{\circ}$, the attention is focused on the rang
of $84^{\circ}$ to $85^{\circ}$. The Kerr rotation can be enhanced
with the increasing of axion angle $\Theta$. Surprisingly, we can
see that the magnitude of the Kerr angle can reach about
$90^{\circ}$ at a fixed incident angle about $84.3^{\circ}$. In
addition, the Kerr angle $\theta_{K}$ reverse its direction when the
axion angle changes the sign so that we can easily manipulate the
Kerr rotation by adjusting the axion angle.

\begin{figure}
\includegraphics[width=8cm]{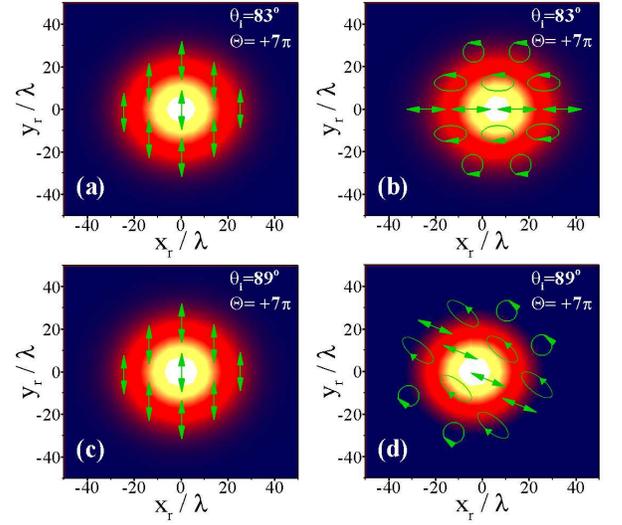}
\caption{\label{Fig3} (Color online) Schematic diagram for electric
field intensity and polarization distribution of light beam. Here
the parameters are $\theta_{i}=83^{\circ}, 89^{\circ}$ and
$\Theta=+7\pi$. (a) and (c) denote the intensity and polarization
distribution without considering spin-orbit coupling electric field
component. The spin-orbit coupling condition are shown in (b) and
(d). The intensity is plotted in normalized units. }
\end{figure}

In fact, the Kerr rotation in TIs is accompanied by the photonic SHE
that is not considered previously. Next we numerically discuss this
effect and reveal the unusual results caused by the axion coupling.
The spin-dependent splitting rotation angle manifesting for the
relationship $\tan\theta_{S}=\delta_{+}^{y}/\delta_{+}^{x}$ is
discussed in the Fig.~\ref{Fig2}(b). It represents some interesting
properties with different axion angles. The results can be explained
by considering the in-plane and transverse spin-dependent splitting,
respectively. First, we analyze the in-plane one. In the past work,
the in-plane spin-dependent splitting can not be observed in the
case of horizontal polarization or vertical
polarization~\cite{Qin2011}. However, in present research, this
effect surprisingly happens in these two polarized states. The
reason is ascribed to the axion coupling effect induced nonzero
cross-polarized reflection coefficients $r_{ps}$ and $r_{sp}$ from
Eq.~(\ref{SRM}). However, at the usual air-glass interface, the
cross-polarized reflection coefficients are equal to zero so that
the in-plane spin-dependent splitting can not be observed.
Remarkably, as we will investigate in the follows, the in-plane
shifts are very sensitive to the value of axion angle. So the
tunable value and sign of the axion angle allow us to describe a
method for modulating the in-plane spin-dependent splitting.

Figure~\ref{Fig2}(c) shows the in-plane shifts in photonic SHE
changing with the incident angle and the axion angle. Here we just
consider the left-circularly polarized component and the polarized
state is chosen as horizontal polarization. For a positive (or
negative) axion anlge ($\Theta\neq0$), the absolute values of the
in-plane shifts first rise with the increasing of incident angle.
Then the displacements decrease after leaving away from the maximum
and reach the negative maximum value. For a fixed incident angle,
the absolute values of the displacements increase when the axion
angle grows ($\Theta=\pm7\pi, \pm15\pi$). When the axion angle
changes its sign, the in-plane shifts reverse its direction too.
Obviously, the in-plane spin-dependent splitting is sensitive to the
variations of axion angle. Therefore we can determine the values of
the axion coupling by detecting the in-plane shifts of photonic SHE
or modulate the in-plane displacments by manipulating the magnitude
of axion angle. It should be noted that, as for the TIs, the
in-plane spin-dependent splitting origin from the axion coupling
effect. If we set the axion angle $\Theta=0$, the Fresnel
coefficients obtaining from Eq.~(\ref{TIRM}) can reduce to the
result of usual media such as BK-7 glass and the in-plane
spin-dependent splitting disappear.

The transverse spin-dependent splitting in photonic SHE is described
in Fig.~\ref{Fig2}(d). We also calculate the transverse shifts
varying with the incident angle $\theta_{i}$ and axion angle
$\Theta$. For a given axion angle, the transverse displacements
firstly rise with the increase of $\theta_{i}$. After reaching the
peak value at the incident angle about $83.6^{\circ}$, the shift
decreases rapidly and then get the negative maximum value. As shown,
the transverse displacements can be tuned to a negative or positive
value by adjusting the incident angle, which represents a switchable
property. However, with the variations of axion angle $\Theta$, the
transverse shifts are almost identical. That is to say, the
transverse spin-dependent splitting is insensitive to the axion
coupling effect. In fact, there really exists tiny differences
between various displacement curves. But these differences are too
small to be distinguished in the picture. It should be noted that,
however, the transverse displacements can represent a significant
variation when the axion angle is chosen as a large value.
\begin{figure}
\includegraphics[width=7.5cm]{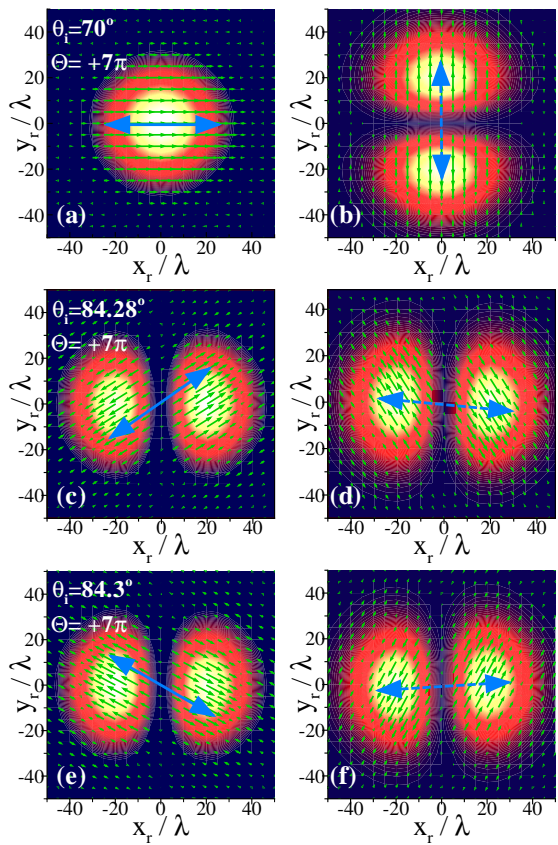}
\caption{\label{Fig4} (Color online) The main electric field and
cross-polarized electric field components of light beam reflected on
TIs interface. Here, the tiny arrows denote the direction of
electric field vector. And the axion angle is chosen as
$\Theta=+7\pi$. (a), (c), and (e) show the main electric field
distribution for different incident angles. The solid arrows
represent the direction of the Kerr rotation. The cross-polarized
electric field components are described in (b), (d), and (f). Here
the dashed arrows stand for the direction of spin-dependent
splitting and the intensity is plotted in normalized units. }
\end{figure}

From Eq.~(\ref{spectrum}), the electric field component
$\exp\left[k_{rx}(\pm i\eta-\vartheta)\right]\exp(\pm
ik_{ry}\delta)$ including the spin-orbit coupling part are used for
field centroid calculation. However, it also affects the
polarization structure of Kerr effect for certain incident angle
ranges. And the other electric field component
$ir_{sp}(\theta_{i})/r_{pp}(\theta_{i})$ takes the great role in
Kerr rotation. Figures~\ref{Fig3}(a)-\ref{Fig3}(d) show the electric
field intensity and polarization distribution of these two
components. Here the incident angle is chosen as $83^{\circ}$ or
$89^{\circ}$ and the axion angle is selected as $+7\pi$. From
Figs.~\ref{Fig3}(a) and~\ref{Fig3}(c), we can see that the
polarization is mainly in vertical direction. When the incident
angle is less than about $84^{\circ}$, the electric field component
$ir_{sp}(\theta_{i})/r_{pp}(\theta_{i})$ represent a tiny value and
the first spin-orbit coupling electric field component can affect
the Kerr rotation. Therefore the Kerr angle shows a small rotation.
As the incident angle increasing, the second component plays the
major role in the Kerr rotation and Kerr angle can reach a huge
value (about $90^{\circ}$). Figures~\ref{Fig3}(b) and~\ref{Fig3}(d)
describe the electric field intensity of the spin-orbit coupling
component. Here, the blue arrows, circles, and ellipses represent
the polarization distribution. When the spin-dependent splitting
occurs, the photons with opposite helicities accumulated at the
opposite edges of the beam.

Generally speaking, the photonic SHE including in-plane and
transverse spin-dependent splitting can be explained by the
cross-polarized effect~\cite{Bliokh2006}. The superposition of the
main electric field and cross-polarized electric field components
finally cause the photons with opposite helicities accumulated at
the opposite edges of the beam, which is the feature for the
spin-dependent splitting. In this work, owing to the axion coupling
effect, the polarization direction of reflected light's central wave
vector changes a tiny value showing a difference with the initial
polarization direction. So the direction of the cross-polarized
electric field vector is also not perpendicular to initial
polarization direction. It is perpendicular to the polarization
direction of reflected light's central wave vector. We note that the
analysis of the cross-polarized effect is useful for studying of
both Kerr rotation and photonic SHE.

According to Eq.~(\ref{FRM}), we can obtain the reflected electric
field by employing the Fourier transformations. Figure~\ref{Fig4}
describes the cross-polarized effect of the light beam reflected on
the TIs interface. Here the calculation parameters are chosen as the
same in Fig.~\ref{Fig2}. Figures~\ref{Fig4}(a),~\ref{Fig4}(c),
and~\ref{Fig4}(e) show the main electric field distribution of
reflected beam. We draw not only the electric field intensity but
also the electric field vector. In the above analysis, we know that
the direction of Kerr rotation can be registered in the polarization
direction of reflected light's central wave vector. From the
picture, we can see that the direction of central electric field
vector (the tiny arrows) undergoes a rotation with different
incident angles. So the Kerr effect also rotates its direction (the
solid arrows). The cross-polarized components are described in the
Figs.~\ref{Fig4}(b),~\ref{Fig4}(d), and~\ref{Fig4}(f). It is noted
that the cross-polarized effect stand for photons with opposite
helicities accumulated at the opposite edges of the beam. Therefore
the direction of cross-polarized electric field intensity
distribution can be regarded as the direction of spin-dependent
splitting (the dashed arrows). We can select the cross-polarized
components by adding a polarizer with the optical axis perpendicular
to the polarization direction of reflected light's central wave
vector. In this condition the shape of cross-polarized electric
field intensity represents a symmetrical double-peak distribution,
which can be seen as a reference. We rotate the polarizer until get
this symmetrical double-peak distribution. Then the Kerr angle can
be obtained from the rotation angle of polarizer.

The axion coupling effect is an important phenomenon, yet it is hard
to be observed experimentally. Some theoretical methods have been
proposed to detect this effect such as measuring the Faraday
rotation angle and the Goos-H\"{a}nchen
displacements~\cite{Maciejko2010,Chang2009}. Here we also
theoretically propose an optical method, the signal enhancement
technique known as weak
measurements~\cite{Aharonov1988,Ritchie1991}, to measure this
topological phenomenon. Note that the weak measurement technique has
attracted a lot of attention and holds great promise for precision
metrology~\cite{Pryde2005,Dixon2009,Dressel2012,Dennis2012}. In the
previous works, the in-plane and transverse spin-dependent splitting
were measured by using this
technique~\cite{Qin2011,Luo2011,Zhou2012b}. We have established the
relationship between axion angle and photonic SHE induced
displacements. If these displacements can be measured, we can
determine the strength of axion coupling. However, under this
condition, the transverse shifts are insensitive to the variations
of axion angle. Therefore we choose to measure the axion coupling by
detecting the in-plane shifts with weak measurements.
\begin{figure}
\includegraphics[width=8cm]{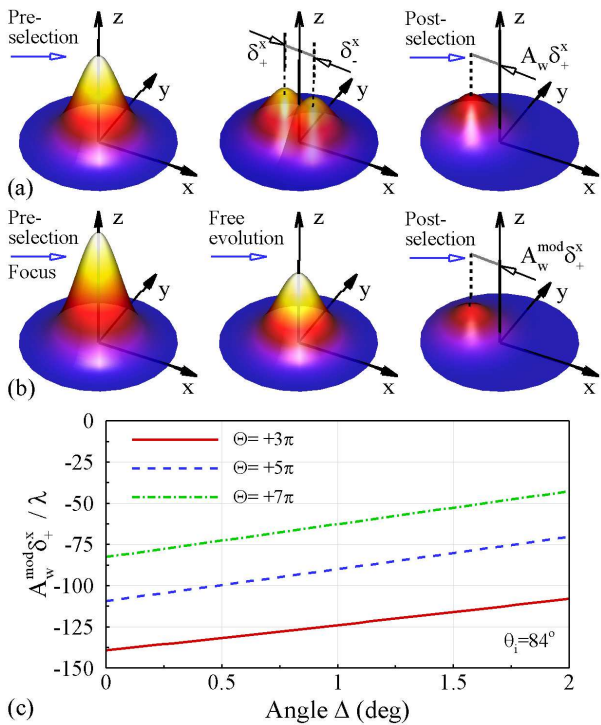}
\caption{\label{Fig5} (Color online) The theory and results of weak
measurements for measuring the axion coupling effect. (a) The two
circularly polarized components interfere destructively after the
second polarizer so that the spin-dependent shifts is significantly
amplified with the amplified factor $A_{w}$. (b) When both of the
pre- and post-selection mechanism and free propagation of light beam
are involved, the final centroid is proportional to $A_{w}^{mod}$
which can be much lager than $A_{w}$. (c) shows the amplified
in-plane displacements changing with the variation of axion angle
$\Theta$ and amplified angle $\Delta$. The incident angle is fixed
to $\theta_{i}=84^{\circ}$ and the observation plane is selected as
$z_{r}=200z_{R}$. Here $z_{R}=k_{0}w_{0}^{2}/2$, $w_{0}$ is the beam
waist. }
\end{figure}

Figure~\ref{Fig5} denotes the weak measurement theory and results
for measuring the in-plane spin-dependent splitting. Here the detail
equipments description and experiment analysis can be found in the
previous works~\cite{Luo2011,Zhou2012a,Zhou2012b}. All of the above
experimental equipments are suitable for the particular wavelength
(for the refractive index n=10 of TIs). In fact, there exists two
types of amplified factors $A_w$ and $F$ similar to that in
Ref.~\cite{Hosten2008}, which gives an enhanced displacement
proportional to initial one $\delta_w=A_{w}^{mod}\delta_+^x$. Here
$A_{w}^{mod}$=$|A_w|F$, $\delta_+^x$ and $\delta_w$ represent the
initial and amplified in-plane shifts. Figure~\ref{Fig5}(a) shows
the pre-selection and post-selection polarization give rise to an
interference in the observation plane, shifting it to its final
centroid position proportional to $A_w=\delta_w/\delta_+^x$. It is
noted that the imaginary $A_{w}$ denotes a shift in momentumspace
corresponding to free evolution of beam. Then the amplified factor
turns out to be $A_{w}^{mod}$=$|A_{w}|F$ which can be much larger
than $A_{w}$. The extra amplified factor $F$ depends on the initial
state of the beam and the degree of its free evolution before the
observer [Fig.~\ref{Fig5}(b)].

The incident beam is focused by the lens and is pre-selected in the
H polarization, and then it undergoes post-selected in the
polarization state with
$\mathbf{V}=\cos\varphi\mathbf{e}_{rx}+\sin\varphi\mathbf{e}_{ry}$.
Here $\varphi=\arctan(r_{sp}/r_{pp})+\pi/2+\Delta$. We note that, in
the previous weak measurement system~\cite{Luo2011}, the two
polarization are nearly perpendicular to each other with an angle of
$90^{\circ}\pm\Delta$. The relevant amplitude of the reflected field
at a plane can be obtained as $\mathbf{V}\cdot\mathbf{E}_{r}$
allowing for calculation of amplified displacement. Here the free
evolution of light beam between the position of initial focused
waist and the position of observer plays great role in the amplified
factor $F$. The theoretical amplified shifts are shown in
Fig.~\ref{Fig5}(c). Here the incident angle is fixed to
$84^{\circ}$. Then we obtain the amplified displacements varying
with axion angle and amplified angle $\Delta$. For a fixed angle
$\Delta$, the amplified in-plane shifts change clearly with the
different axion angles ($\Theta=+3\pi, +5\pi$, and $+7\pi$). So we
can measure the axion coupling effect by determining the in-plane
displacements with weak measurements.

\section{Conclusions}
In conclusion, we have examined the photonic spin Hall effect (SHE)
of light beam reflected from the air-topological insulators (TIs)
interface due to axion coupling. We have found that the spin-orbit
coupling in photonic SHE can be routed by adjusting the strength of
the axion coupling. The in-plane spin-dependent splitting is
sensitive to the axion angle variations of TIs so that the enhanced
and switchable in-plane shifts was observed with different axion
coupling. However, the transverse spin-dependent splitting is
insensitive to the axion coupling in the case of axion angle with
small value. We have also found that, by adjusting the axion angle,
the magnitude and direction of Kerr rotation angle can be modulated.
In addition, the field structure for the reflected beam has been
analyzed. The cross-polarized effect induced symmetrical double-peak
field structure can be used to determine the Kerr rotation.
Importantly, a signal enhancement technique known as weak
measurements has been theoretically proposed to measure this
topological phenomenon. These findings offer us the potential
methods for determining the strength of the axion coupling and
provide new insight into the interaction of light with TIs.

\begin{acknowledgements}
This research was partially supported by the National Natural
Science Foundation of China (Grants Nos. 61025024 and 11274106).
\end{acknowledgements}

\end{document}